\documentclass[aps,prc,twocolumn,groupedaddress,showpacs,floatfix,nofootinbib]{revtex4}
\usepackage{amssymb}
\usepackage{graphicx}
\usepackage{bm}

\newcommand{\be}{\begin{equation}}
\newcommand{\ee}{\end{equation}}

\begin{document}

\title{Dark matter annihilation and jet quenching phenomena in the early universe}

\author{Igor 
N.~Mishustin}

\affiliation{Frankfurt Institute for Advanced Studies,
D--60438 Frankfurt am Main, Germany}
\affiliation{National Research Center ''Kurchatov Institute'',\\
 123182 Moscow, Russia}

\begin{abstract}

Dark-matter particles like neutralinos should decouple from the hot cosmic plasma at temperatures of about 40 GeV. Later they can annihilate each other into standard-model particles, which are injected into the dense primordial plasma and quickly loose energy. This process is similar to jet quenching in ultrarelativistic heavy-ion collisions, actively studied in RHIC and LHC experiments. Using empirical information from heavy-ion experiments I show that the cosmological (anti)quark and gluon jets are damped very quickly until the plasma remains in the deconfined phase. The charged hadron and lepton jets are strongly damped until the recombination of electrons and protons. The consequences of energy transfer by the annihilation products to the cosmic matter are discussed.

\end{abstract}


\maketitle


{\bf \it Introduction.}
Presently the hypothesis on Dark Matter (DM) existence in our universe is rather popular and broadly discussed in the scientific community. In particular, it helps to explain the fluctuation spectrum of the Cosmic Microwave Background (CMB) radiation measured with high presision by WMAP and PLANCK space missions. In this paper I will investigate possible DM manifestations in the early universe. Following many authors, see e.g. \cite{LW, KT}, I assume that the DM is made of Weakly Interacting Massive Particles (WIMPs), which could be the 
lightest SUSY particle e. g. neutralino $\chi$. Since the mass of this hypothetical particle is still poorly constrained, for numerical estimates below I will use $m_\chi$=1 TeV. The neutralinos are supposed to be Majorana fermions so that they may annihilate each other in the binary collisions,
\be
\chi\chi\rightarrow l\bar{l}, q\bar{q}, gg, \gamma\gamma,...                   \label{annih}
\ee
where the r. h. s. contains the Standard Model (SM) particles: (anti)leptons, (anti)quarks, gluons, photons etc.
It is believed that at early stages of the cosmic evolution neutralinos were in statistical equilibrium with the primordial plasma made of SM particles due to the balance berween the creation and annihilation reactions.

{\bf \it Main equations.}
The rate equation governing the neutralino number density $n$ can be written as \cite{LW, KT}\footnote{The units with c=$\bar{h}$=1 are used below.} 
\be
\frac{1}{V}\frac{dN}{dt}=\frac{dn}{dt}+3Hn=-\langle\sigma_Av\rangle\left(n^2-n^2_{\rm eq}\right),       \label{dn}
\ee
where $\sigma_A$ is the total annihilation cross section, $v$ is the relative velocity and $H=\dot{a}/a$ is the Hubble parameter controlling the expansion rate ($V\propto a^3$). According to the Friedman equation it is expressed as
\be
H^2=\left(\frac{\dot{a}}{a}\right)^2=\frac{8\pi G_N}{3}\rho=\frac{2}{3M_{\rm Pl}^2}\rho,   \label{H}
\ee
where $M_{\rm Pl}==1/\sqrt{4\pi G_N}=3.38\cdot 10^{18}$ GeV is the reduced Planck mass.
The first term in the r. h. s. of eq. (\ref{dn}) corresponds to the loss of $\chi$ particles due to annihilation, while the second term approximately accounts for the regeneration reactions. The equilibrium density of $\chi$ particles at $T\ll m_{\chi}$ is given by
\be
n_{\rm eq}=\nu\left(\frac{m_\chi T}{2\pi}\right)^{3/2}\exp\left(-\frac{m_\chi}{T}\right),       \label{neq}
\ee
where $\nu$=2 is their spin-degeneracy factor. 
Equation (\ref{dn}) should be solved together with the thermodynamic relation expressing the total energy conservation:
\be
\frac{d\rho}{dt}+3H(\rho+p)=0,                                              \label{drho}
\ee
where $\rho$ and $p$ are the total energy density and pressure of the system. 
I assume that they  can be represented as the sum of two components: the "radiation" component, which includes all relativistic SM degrees of freedom, and nonrelativistic "cold" DM component, represented by $\chi$ particles, i.e.
\be
\rho=\epsilon+\left(m+\frac{3}{2}T\right)n, ~~~~p=\frac{\epsilon}{3}+nT~.    \label{EOS}
\ee
The energy density $\epsilon$ and entropy density $s$ of the radiation are 
\be
\epsilon=\nu_{*}\frac{\pi^2}{30}T^4,~~~~~s=\frac{4\epsilon}{3T}=\nu_{*}\frac{2\pi^2}{45}T^3, \label{en}
\ee
where $\nu_{*}$ is the effective number of relativistic degrees of freedom which depends on temperature. At T$\sim$50 GeV, $\nu_{*}$ is close to 90.
During the radiation-dominated era, when $\epsilon\gg mn$, the scale factor varies as $\sqrt{t}$ and the Hubble parameter (\ref{H}) can be expressed as
\be
H=\frac{1}{2t}=\frac{H_0}{x^2}, ~~~~H_0=\frac{\pi}{3\sqrt{5}}\nu_{*}^{1/2}\frac{m_{\chi}^2}{M_{\rm Pl}},        \label{Tvst}
\ee
where new variable $x=m/T$ has been introduced to be used below instead of t. 
Also, it is convenient to consider the quantity $Y= n/s$, which would stay constant in case of the conserved entropy and $\chi$-particle number. 

Now the equation (\ref{dn}) can be rewritten as:
\be
\frac{dY}{dx}= -\frac{\lambda}{x^{2-k}}\left(1+xY\right)\left(Y^2-Y_{\bf eq}^2\right),          \label{dY}
\ee
where $Y_{eq}=bx^{3/2}e^{-x}$, $b=0.145\frac{\nu}{\nu_{*}}$ and
\be
\lambda=\frac{2\pi}{3\sqrt{5}}\nu_{*}^{1/2}\sigma_0M_{\rm Pl}m_{\chi}.         \label{lambda}
\ee
In eq. (\ref{dY}) the parametrization $\langle\sigma_Av\rangle=\sigma_0/x^k$ was introduced, which allows to consider the $\chi\chi$ annihilation in s-wave (k=0) and p-wave (k=1) states. One can find the detailed discussion of this equations e. g. in ref. \cite{KT}. The additional factor $(1+xY)$ in eq. (\ref{dY}) comes from the back reaction of the $\chi\chi$ annihilation on the plasma entropy.

{\bf \it Decoupling of DM particles.}
At early times the annihilation and regeneration reactions are very fast so that SM and DM particles are in thermodynamical equilibrium. Since neutralinos are supposed to be weakly-interacting particles, their annihilation cross section should be small. 
As was first shown in ref. \cite{LW}, the asymptotic abundance of such particles is determined almst entirely by the parameter $\sigma_0$. To get the present DM energy density $\Omega_d\approx 0.27$ and assuming s-wave annihilation (k=0) one needs $\sigma_0=\approx 3\cdot 10^{-26}$ cm$^3$s$^{-1}$= $10^{-10} {\rm fm}^{2}$ \cite{LW,Salati}. This value will be used in the numerical estimates below. 
When the system cools down sufficiently (the density of $\chi$-particles drops), the creation of heavy $\chi\chi$-pairs becomes inefficient and they fall out of equilibrium with thermal bath. The decoupling (freeze-out) temperature $T_{\rm f}$ is determined from the transcendental equation (see details in \cite{KT})
\be
x_{\rm f}=\ln(\left[(k+1)b\lambda\right]-\left(k+\frac{1}{2}\right)\ln(x_{\rm f}).               \label{xf}
\ee
where $x_{\rm f}=m_\chi/T_{\rm f}$ and $\lambda$ is defined in eq. (\ref{lambda}).
With parameters specified above we get $\lambda =8\cdot 10^{13}$ leading to $x_{\rm f}\approx$25 and 
$T_{\rm f}\approx$40 GeV.  
After freeze-out the regeneration term in eq. (\ref{dY}) drops rapidly and the abundance of neutralinos is changing mainly due to their mutual annihilation. 
Then the function $Y(x)$ can be easily found by neglecting $Y_{\rm eq}$ term in the r. h. s. of eq. (\ref{dY}),
\be
\frac{1}{Y}=\frac{1}{Y_{\rm f}}+\frac{\lambda}{k+1} \left(\frac{1}{x^{k+1}_{\rm f}}-\frac{1}{x^{k+1}}\right).    \label{1/Y}
\ee
The $Y$ values at freeze-out ($x=x_{\rm f}$) and at asymptotically late times ($x\gg x_{\rm f}$) can be expresed as
\be
 Y_{\rm f}=\frac{x_{\rm f}^{k+2}}{(k+1)\lambda},~~~~~~ Y_{\infty}=\frac{(k+1)x_{\rm f}^{k+1}}{\lambda}.  \label{Y(x)}
\ee
The ratio of energy densities of nonrelativistic DM particles ($mn$) and relativistic plasma ($\epsilon=\frac{4}{3}Ts$) can be written as  
\be
\eta_{DM}=\frac{m_{\chi}n}{\epsilon}=\frac{4}{3}\frac{m_{\chi}n}{Ts}=\frac{4}{3}xY.
\ee
It is interesting to note that at late times this ratio depends only weakly (logarithmically) on the DM particle mass $m_{\chi}$, because $Y$ is inversely proportional to $\lambda\propto m_\chi$, see eq. (\ref{lambda}). Thus it is determined mainly by the annihilation cross  section $\sigma_0$. The $Y$ values at freeze-out ($T\approx$40 GeV) and at the radiation-matter (rm) equality time when $\eta_{DM}\approx 1$ ($T\approx$ 0.8 eV for $k=0$) are, respectively, 
$Y_{\rm f}\approx 8\cdot 10^{-12}$ and $Y_{\rm rm}=\approx 3\cdot 10^{-13}$. In case of k=1, $Y_{\rm rm}$ acquires an extra factor $2x_{\rm f}\approx 50$. Then, to have the same mass density of dark matter today we should choose larger annihilation cross section, $\sigma_0\approx 5\cdot 10^{-9}$ fm$^{2}$. 

From this analysis we conclude that the abundances of DM particles at freeze-out and rm-equality time differ by a large factor, $\frac{x_{\rm f}}{(k+1)^2}\approx 25\div6$ for k=0 and k=1, respectively. This means that there was much more DM particles per entropy unit at freeze-out than at present. At later times they have destroyed each other in the annihilation reactions (\ref{annih}). In principle, the stable annihilation products, such as $p$, $\bar{p}$,  $e^{+}$, $e^{-}$, $\gamma$, $\nu$, $\tilde{\nu}$ with energies in the range of $m\chi \sim 1$ TeV, may survive until present time. The "excess" in (anti)protons and positrons with energies 300$\div$1000 GeV is indeed observed in cosmic rays by several collaborations, e.g. PAMELA \cite{PAMELA} and AMS-02 \cite{AMS-02}. However, at early times the DM annihilation  products will be a subject to the interaction with the primordial plasma and, therefore, will be strongly quenched.

{\bf \it Cosmological jet quenching.}
After decoupling of dark matter, the SM particles produced in its annihilation, see eq. (\ref{annih}), are injected into the hot and dense primordial plasma. At temperatures around and below 40 GeV it  contains photons, gluons,  5 lightest flavors of quarks and antiquarks, and all species of leptons, that gives $\nu_{*}\approx 80$. The interaction of high-energy $q$ and $\bar{q}$ propagating through this plasma can be analysed within standard approaches used previously for the description of quark-jet energy loss in relativistic heavy-ion collisions, see ref. \cite{Bass} and references therein. The temperature of the quark-gluon plasma produced in such collisions reaches values of about 1 GeV, resulting in parton densities of the order of 100 fm$^{-3}$ for $N_{\rm f}$=3. This is still much smaller than the parton densities in the cosmic plasma, which may reach the values $10^7$ fm$^{-3}$ at $N_{\rm f}$=5 and $T$=40 GeV. 

Since neutralinos are very heavy ($m_\chi \gg T$), they can be considered at rest. On the other hand, the partons produced in the annihilation process $\chi\chi\rightarrow q\bar{q}$ are ultrarelativistic, their initial energies are $E_0\approx m_\chi=1$ TeV. As follows from the calculations, the stopping time of such partons is always much shorter than the characteristic expansion time of the universe, $t_{H}=1/{H}$. Assuming that the partons produced at time $t_0$ move through the uniform plasma with the speed of light, one can rewrite the standard equation for radiative energy losses \cite{BDMPS} as 
\be
\frac{dE}{dt}=\frac{\alpha_s(T_0)N_c}{12}\hat{q}(T_c)\left[\frac{\nu_{p}(T_0)T_0^3}{\nu_{p}(T_c)T_c^3}\right](t-t_0),   \label{qhat}
\ee
where $T_0$ is the plasma temperature at $t=t_0$, $\alpha_s(T)$ is the temperature-dependent strong coupling constant, $\nu_{p}(T)$ is the degeneracy factor for partonic plasma which also changes with temperature. The jet quenching parameter $\hat{q}$ is proportional to the parton density in primordial plasma. In the above expression it is normalized to the value $\hat{q}(T_c)$ at the critical temperature for the deconfinement phase transition $T_c\approx 170$ MeV. Extrapolating the fits of experimental data from ref. \cite{Qin} to this temperature gives $\hat{q}(T_{\rm c})/T_{\rm c}^3\approx 7$.  

Equatiion  (\ref{qhat}) allows to estimate the time interval required to quench the parton, 
$\Delta E\approx m_{\chi}$, i. e.
\be
\Delta t=\frac{C_1}{T_0}\left(\frac{m_\chi}{\alpha(T_0)N_cT_0}\right)^{1/2}\left[\frac{\nu_p(T_c)}{\nu_p(T_0)}\right]^{1/2},~~~~~~~~~C_1\approx 1.8 \label{Delta}
\ee
The corresponding damping length, $c\Delta t$ varies from $0.1$ fm at $T_0=T_{\rm f}=40$ GeV ($N_f$=5) to about 230 fm at $T_0\approx T_c=170$ MeV ($N_f$=3). The first value looks unrealistically short.  
It is difficult to imagine how a TeV parton could loose its full energy within a fraction of fm! Formally, this follows from the extremely high energy density in the cosmic plasma, reaching values of about $10^{10}$ GeV/fm$^{3}$ at $T=40$ GeV.  Perhaps, some additional screening mechanisms, beyond the LPM effect included in eq. (\ref{qhat}), should be considered. It is interesting to note that the maximum stopping distance of light quarks calculated within the gauge/gravity duality for $\cal{N}$=4 SYM strongly-coupled plasma is expressed as \cite{SYM}
\be
\Delta z=\frac{C_2}{T}\left(\frac{m_\chi}{\sqrt{\lambda}T}\right)^{1/3},~~~~~~~    C_2=.526,          \label{SYM}
\ee
where $\lambda=g^2_{YM}N_c$ is the t'Hooft coupling constant, which is fixed to 1 in numerical estimates below. This formula has the same structure as eq. (\ref{Delta}), but the power is 1/3 instead of 1/2. For temperatures $T=40$ GeV and 170 MeV  the predicted stopping distances are 0.008 fm and 11.2 fm, respectively. They are even shorter than the values obteined with the pQCD-based calculation! Definitely, more detailed studies of this issue are required in the future. 

The "strong-damping" regime continues
until the injection temperature drops below the critical temperature $T_c$ when free color charges disappear. At lower temperatures the energy loss of high-energy quarks and antiquarks is determined by the interaction with hadronic species. Some information about such  interactions has been obtained from deep-inelastic scattering of electrons off cold nuclei, when a fast quark or antiquark can be produced inside the cold nucleus. As follows from the analysis of ref. \cite{Qin}, the corresponding value of the transport coefficient $\hat{q}$ in this case is about 30 times smaller than in the QGP. But this is more than sufficient to quench the products of DM decay within a microscopic scale.

The hadrons other than nucleons practically disappear from the cosmic plasma at temperatures below 50 MeV ($t>0.4$ ms). At $T=1\div50$ MeV the baryon to photon ratio $\eta_B$ practically does not change  and is about $1.7\cdot 10^{-9}$ \cite{SM}. The baryon (nucleon) density at this stage can be calculated as $n_N=\eta_B n_\gamma$, that gives approximately $6\cdot 10^{-12}$ fm$^{-3}$ at $T$=50 MeV and $5\cdot 10^{-17}$ fm$^{-3}$ at $T$=1 MeV. The mean-free path of hadrons from the DM decay in such a medium can be estimated as $n_N\sigma_{hN}$, where $\sigma_{hN}\approx 100$ mb. In the considered temperature interval it changes from 15 $\mu m$ to about 1 m. Since in each inelastic collision the leading particle looses about half of its c.m. energy, it will require about 5-6 collisions to quench the hadronic decay products.

At even later stage only electron-positron pairs, neutrinos and photons remain in the plasma with very small, $~10^{-9}$, admixture of baryons. The electromagnetic energy losses for charged decay products can be estimated from the expression (v$\approx$ c) \cite{Thoma}
\be
\frac{dE}{dt}=\frac{e^4T^2}{24\pi}\left(\ln\frac{E}{M}+C\right),~~~~~~~C\approx  2.6,
\ee
where M is the charged particle mass, $e=\sqrt{4\pi\alpha}\approx0.3$ is the electron charge. At injection temperatures $T_0\sim 1$ MeV the estimated damping range for protons/antiprotons and electrons/positrons is about 1 mm. The EM energy losses remain important practically until the e-p recombination is over ($T\sim$0.3 eV), when the corresponding range is about 10$^6$ km. At later times the nonrelativistic matter dominates over radiation, the universe expands even faster ($a\propto t^{2/3})$, and DM decay products propagate almost freely over cosmological distances.

Especially interesting are the $\chi\chi$ annihilation channels containing electrically-neutral SM particles such as neutrons/antineutrons ($n\bar{n}$), neutrinos/antineutrinos ($\nu\tilde{\nu}$) and photons ($\gamma \gamma)$. They should decouple from the primordial plasma at earlier times and thus may bring a stronger signal of the dark matter annihilation. 
The life time of (anti)neutrons is rather long, about 15 min times a Lorentz-factor of order 10$^3$, Therefore, they will decay already after the recombination era. At this stage the EM energy losses of their decay products ($p$, $\bar{p}$, $e^{-}$, $e^{+}$) are very small. Therefore, they may survive until present time and contribute to the observed fluxes of cosmic rays. Indeed, a significant excess of high-energy antiprotons and positrons are observed by several space-based experiments, see refs. \cite{PAMELA, Planck, AMS-02}. It is interesting that using data presented in ref. \cite{Ting}, one can see that the fluxes of antiprotons and positrons with energies around 300 GeV are equal within a factor of 2. 
  
Perhaps, the most promising signal of DM presence in the universe can be provided by high-energy (anti)neutrinos. Because of very small interaction cross section with ordinary matter they should decouple from the primordial plasma at earlier times than hadrons, and therefore may be present in cosmic rays today in a greater amount. However, the calculation of their present flux is not a trivial task, because the direct production channel $\chi\chi\rightarrow 
\nu\tilde{\nu}$ may be suppressed by a very small neutrino mass. Instead one may consider indirect decay channels involving heavy particles like $\tau^{+}\tau^{-}$ or $W^{+}W^{-}$, which then decay to (anti)neutrinos, see e. g. ref. \cite{Allahverdi}. An interesting possibility is that the residual DM particles are gravitationally trapped inside the stars like Sun or in the galactic halo. Then the high-energy (anti)neutrinos from the DM annihilation could in principle be registered by large-scale terrestrial detectors like IceCube \cite{IceCube} and Baikal-GVD \cite{Baikal}. Some interesting events have been already reported \cite{Aartsen}.


{\it Effects on the primordial cosmic matter.} From the above analysis one can conclude that all charged decay products of dark matter will be strongly damped in the primordial plasma until the recombination era is over. This means that their energy will be transferred to the plasma leading to its increasing entropy. The response of the medium to the energy deposition by energetic partons produced in relativistic heavy-ion collisions was considered recently by many authors, see e. g. refs. \cite{Solana, Chaudhuri, Betz1, Bass}. 
In particular, the collective excitations of the medium in the form of Mach cones and diffusion wakes were studied in detail in ref. \cite{Betz2}. As follows from the calculations, in case of two back-to-back partons the deposited energy is confined in a region inside the two oppositly-moving Mach cones. The transverse size of this domain is about 2$c_s$t and longitudinal size is about 2ct, where $c_s\approx c/\sqrt{3}$ is the speed of sound. 

An interesting result of ref. \cite{Betz2} is that a hot spot (diffusion wake) is formed behind each of two receding partons. They survives for a long time,  even after the partons are fully quanched (see corresponding 2d plots in ref. \cite{Betz2}). Therefore, the fast partons produced from the neutralino annihilation will continuously generate strong perturbations in the plasma with the characteristic scale of about $c_s\tau$, where $\tau$ is the life time of such perturbations. It should depend on temperature and thermal conductivity of the plasma. Such a system looks like a sparkling liquid where each spark is accompanied by a supersonic boom. 

Strong perturbations  discussed above may lead to new interesting phenomena. For instance, one may expect the formation of a mixed quar-hadron phase even when the equilibrium matter has a smooth crossover-type phase transition. Indeed, when the temperature of the plasma drops below the pseudo-critical value for the deconfinement phase transition, $T_c\approx$ 170 MeV, the hot spots generated by the annihilation partons may still remain in the deconfined phase. Moreover, such states may be formed even at temperatures as low as 50 MeV, when hadrons are still abundent in the primordial plasma.   

It has been realized already a long time ago, see e. g. \cite{Ellis}, that the high-energy DM annihilation products may significantly modify the BBN predictions at temperatures below 1 MeV. Different nuclear processes induced by these particles are discussed in refs. \cite{Salati, Pospelov}. For instance, high-energy electrons, positrons and photons can induce EM disintegration of D and $^4$He nuclei. Also, fast hadrons may induce spallation reactions on $^4$He leading to fast $^3$H and $^3$He ions. They in turn can initiate endothermic reactions $^3$H+$^4$He$\rightarrow$ $^6$Li+n and $^3$He+$^4$He$\rightarrow$ $^6$Li+p. But the reaction 
p+$^7$Li$\rightarrow$$^4$He+$^4$He should lead to the depletion of $^7$Li. As a result, one should expect more $^6$Li and less $^7$Li produced.    

It should be also noted that the energy deposition by the DM annihilation products may also be very important after the recombination transition. Indeed, the fast charged particles like $p$. $\bar{p}$, e$^{+}$, e$^{-}$ will induce ionization and excitation of the H and He atoms on the way through the cosmic matter. At this stage the densities of H and He atoms are of order 10$^3$ cm$^{-3}$. Using the empirical information about the ionization energy loss of protons in H$_2$ gas, ~5 MeV$\cdot$cm$^2$/g \cite{PDG}, one can evaluate the stoping distance for 1 TeV protons/antiprotons in the cosmic medium on the level of 100 Mpc.  Taking into account that the energy transferred to the atom and $\delta$-electrons is in average of about 30$\div$40  eV, one can estimate the total number of ions produced as ~10$^{10}$!. This may lead to signifiant observable effects in the anisotropy of microwave background radiation, see e. g. \cite{Galli}. Extremely precise measurements of the PLANCK collaboration \cite{Planck} open the possibility to constrain the DM annihilation at this epoch.

{\it Conclusions.} If dark matter is made of wekely-interacting massive particles like neutralinos, their annihilation will lead to interesting phenomena in the early universe. The standard model particles from DM annihilation 
will be injected into the dense primordial plasma and loose energy, similarly to jet quanching phenomena in relativistic heavy-ion colllisions studied at RHIC and LHC. Using the empirical information from these experiments one can estimate damping range of annihilation products at different stages of the universe evolution. The stochastic energy deposition into the primordial plasma should lead to strong nonstatistical fluctuations of its temperature and composition.

 {\it Acknowledgements.} This work was partially support by the Helmholtz International Center for FAIR (Germany) and grant NSH-932.2014.2 of the Russian Ministry of Education and Science.

\end{document}